\documentclass[aps,showpacs,twocolumn,amsmath,amssymb,superscriptaddress,prl]{revtex4-1}

\usepackage[dvipdfmx]{graphicx}
\usepackage[dvipdfmx]{color}

\usepackage{bm}
\usepackage[pagewise]{lineno}

\begin{document}

\title{Surface Rashba-Edelstein Spin-Orbit Torque Revealed by Molecular Self-Assembly}

\author{Satoshi Haku}
\affiliation{Department of Applied Physics and Physico-Informatics, Keio University, Yokohama 223-8522, Japan}

\author{Atsushi Ishikawa}
\affiliation{Center for Green Research on Energy and Environmental Materials, National Institute for Materials Science, Tsukuba, 305-0044, Japan
}

\author{Akira Musha}
\affiliation{Department of Applied Physics and Physico-Informatics, Keio University, Yokohama 223-8522, Japan}

\author{Hiroyasu Nakayama}
\affiliation{International Center for Young Scientists, National Institute for Materials Science, Tsukuba, 305-0047, Japan}

\author{Takashi Yamamoto}
\affiliation{Department of Chemistry, Keio University, Yokohama 223-8522, Japan
}

\author{Kazuya Ando}
\email{ando@appi.keio.ac.jp}
\affiliation{Department of Applied Physics and Physico-Informatics, Keio University, Yokohama 223-8522, Japan}
\affiliation{Keio Institute of Pure and Applied Science (KiPAS), Keio University, Yokohama 223-8522, Japan}
\affiliation{Center for Spintronics Research Network (CSRN), Keio University, Yokohama 223-8522, Japan}

\date{\today}

\begin{abstract}
We report the observation of a spin-orbit torque (SOT) originating from the surface Rashba-Edelstein effect. We found that the SOT in a prototypical spin-orbitronic system, a Pt/Co bilayer, can be manipulated by molecular self-assembly on the Pt surface. This evidences that the Rashba spin-orbit coupling at the Pt surface generates a sizable SOT, which has been hidden by the strong bulk and interface spin-orbit coupling. We show that the molecular tuning of the surface Rashba-Edelstein SOT is consistent with density functional theory calculations. These results illustrate the crucial role of the surface spin-orbit coupling in the SOT generation, which alters the landscape of metallic spin-orbitronic devices.
\end{abstract}

\maketitle

\section{I. introduction}

The spin-orbit coupling is central to magnetism and spintronics~\cite{RevModPhys.89.025008,RevModPhys.89.025006,RevModPhys.87.1213,0034-4885-78-12-124501,RevModPhys.91.035004,hoffmann2013spin}. It describes the relativistic interaction between the electrons' spin and momentum degrees of freedom, which has a dramatic impact on both the equilibrium and nonequilibrium properties of condensed matter. Of particular recent interest is the interplay between the spin-orbit coupling and space inversion symmetry, which triggers a variety of phenomena, such as chiral spin textures and spin-momentum locking~\cite{manchon2015new,soumyanarayanan2016emergent}. In systems with broken inversion symmetry, the spin-orbit coupling lifts the electron-spin degeneracy~\cite{rashba1960properties}. This phenomenon, the Rashba effect, has been observed across various surfaces and interfaces~\cite{PhysRevLett.78.1335,Ast:186807,PhysRevLett.77.3419,PhysRevLett.108.066804}.

The Rashba effect couples the spin and charge transport through the spin-momentum locking; an in-plane charge current induces a transverse spin accumulation~\cite{edelstein1990spin,bel2008magneto}. This phenomenon, the Rashba-Edelstein effect or the inverse spin galvanic effect, can generate spin-orbit torques (SOTs), which allow electric manipulation of magnetization in metallic heterostructures~\cite{miron2011perpendicular,liu2012spinScience,yu2014switching,kurebayashi2014antidamping,fukami2016magnetization}. A prototypical spin-orbitronic system is a Pt/Co bilayer, where the strong spin-orbit coupling of the most fundamental charge-spin converter, Pt, plays a key role in the generation of the SOT. The SOT in such spin-orbitronic devices is expected to be generated by the surface, bulk, and interface spin-orbit coupling. However, the SOT has been generally attributed to two mechanisms: the interface Rashba-Edelstein effect and bulk spin Hall effect~\cite{Gambardella3175,RevModPhys.91.035004,doi:10.1063/1.5041793}. Despite the extensive studies on the SOT for the past decade, the SOT originating from the surface Rashba-Edelstein effect has  been neglected.

In this paper, we report the observation of the SOT arising from the surface Rashba-Edelstein effect: the surface Rashba-Edelstein SOT. The crucial evidence was obtained by using the molecular tuning of the Rashba-Edelstein effect~\cite{nakayama2018molecular}. Using the molecular tuning as a tool to study the SOT, we found that the damping-like (DL) SOT in a Pt/Co bilayer is manipulated by decorating the Pt surface with self-assembled organic monolayers, while the field-like (FL) SOT is unaffected by the molecular self-assembly. This result is consistent with the prediction of the SOT arising from the surface Rashba-Edelstein effect and spin-transfer mechanism, illustrating the crucial role of the surface spin-orbit coupling in spin-orbitronic devices.

\begin{figure}[tb]
\includegraphics[scale=1]{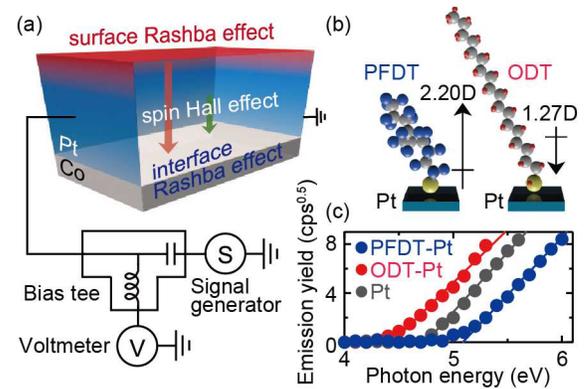}
\caption{
(a) A schematic illustration of the Pt/Co bilayer and experimental setup. (b) A schematic illustration of ODT and PFDT molecules on the Pt surface. The arrows represent the dipole moment of the SAM-forming molecules~\cite{nakayama2018molecular}. (c) Square root of photoelectron emission yield as a function of scan energy measured with an atmospheric photoelectron spectrometer for the pristine Pt/Co (black), ODT-Pt/Co (red), and PFDT-Pt/Co (blue). The solid lines are linear fits, from which the baseline intercept gives the work function. 
}
\label{fig1} 
\end{figure}

\begin{figure*}[tb]
\includegraphics[scale=1]{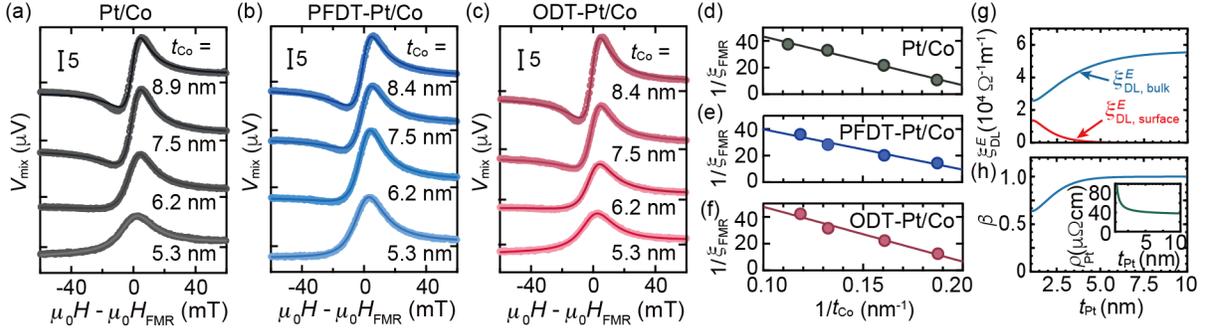}
\caption{
The ST-FMR spectra for the (a) Pt/Co, (b) PFDT-Pt/Co, and (c) ODT-Pt/Co films at $f=10.3$ GHz, where $t_\text{Co}$ is the thickness of the Co layer. The open circles are the experimental data and the solid curves are the fitting results using the sum of the symmetric and antisymmetric functions. 1/$t_\text{Co}$ dependence of $1/\xi_\text{FMR}$ for the (d) Pt/Co, (e) PFDT-Pt/Co, and (f) ODT-Pt/Co films. The solid circles are the experimental data. The solid lines are the linear fitting results using Eq.~(\ref{thicknesseq}). (g) Pt-layer thickness $t_\mathrm{Pt}$ dependence of $\xi_\text{DL,surface}^E$ and $\xi_\text{DL,bulk}^E$, calculated using Eqs.~(\ref{xiEbulk}) and (\ref{xiEsurface}), respectively, for Pt/Co films. (h) $t_\mathrm{Pt}$ dependence of $\beta=\xi_\text{DL,bulk}^E/(\xi_\text{DL,bulk}^E+\xi_\text{DL,surface}^E)$. The inset shows $t_\mathrm{Pt}$ dependence of the Pt-layer resistivity $\rho_\mathrm{Pt}$ calculated using a model that takes into account the carrier scatterings at film surfaces and grain boundaries with the parameters $p = 0.97$, $\rho_{\infty}=31\ \mu\Omega\ \rm cm$, $\lambda= 10$ nm, and $\zeta=0.25$ [for details, see Ref.~\cite{doi:10.1063/1.1644906}]. 
}
\label{fig2} 
\end{figure*}

\section{II. EXPERIMENTAL METHODS}

The molecular tuning of the surface Rashba-Edelstein effect is demonstrated by measuring the SOTs in Pt/Co bilayers using the spin-torque ferromagnetic resonance (ST-FMR). Figure~\ref{fig1}(a) shows a schematic illustration of the device structure. The bilayers were fabricated by radio frequency (RF) magnetron sputtering. Firstly, a Co layer with thickness of $t_\text{Co}$ was grown on a thermally oxidized Si substrate. Then, on the top of the Co layer, a 1-nm-thick Pt layer was sputtered without breaking the vacuum. To study the effect of the decoration of organic monolayers on the surface of the Pt layer, we used 2 mM solutions 1-octadecanethiol (ODT) and 1{\it H},1{\it H},2{\it H},2{\it H}-perfluorodecanethiol (PFDT) in ethanol. The Pt/Co films were immersed into the solution for 20  hours at room temperature, which results in the formation of self-assembled monolayers (SAMs) on the surface of the ultrathin Pt layer [see Fig.~\ref{fig1}(b)]~\cite{onclin2005engineering,de2005tuning,fendler2001chemical,ma2010interface,love2005self,ulman1996formation}. The SAM formation is confirmed by the change of the work function $\varPhi$. Figure~\ref{fig1}(c) shows that the ODT formation decreases $\varPhi$, whereas the PFDT formation increases $\varPhi$. This result is consistent with the opposite dipoles of ODT and PFDT molecules, since the change in the metal work function is associated with the dipole moment of the SAM-forming molecule perpendicular to the surface~\cite{de2005tuning}. 
The SAM formation was also confirmed by x-ray photoelectron spectroscopy~\cite{Supplemental}.

For the ST-FMR measurement, the Pt/Co films were patterned into rectangular strips with 10-$\mu$m width and 100-$\mu$m length by photolithography and Ar ion etching. Along the longitudinal direction of the Pt/Co device, an RF current with a frequency of $f$ was applied and an in-plane external field $H$ was applied at an angle of $45^\circ$ with respect to the longitudinal direction of the device. In the Pt/Co bilayer, the RF current flowing in the Pt layer generates DL- and FL-SOTs, as well as an Oersted field, which excite magnetization precession in the Co layer under the FMR condition: $(2\pi f/\gamma) = \sqrt{\mu_{0}H_\text{FMR}(\mu_{0}H_\text{FMR} + \mu_{0}M_\text{eff})}$, where $\gamma$ is the gyromagnetic ratio, $H_\text{FMR}$ is the FMR field, and $M_\text{eff}$ is the effective demagnetization field~\cite{PhysRevB.92.064426}. The magnetization precession generates a direct-current (DC) voltage $V_\text{mix}$ through the mixing of the RF charge current and oscillating resistance due to the anisotropic magnetoresistance (AMR) in the Co layer~\cite{liu2011spin,zhang2015role}:
\begin{multline}
V_\text{mix}=V_\text{sym}\frac{{W^2}}{{(\mu_0H-\mu_0H_\text{FMR})^2+W^2}} \\
+V_\text{anti}\frac{{W(\mu_0H-\mu_0H_\text{FMR})}}{{(\mu_0H-\mu_0H_\text{FMR})^2+W^2}}, \label{Vmix}
\end{multline}
where $W$ is the spectral width. In the ST-FMR signal, the magnitude of the symmetric component $V_\text{sym}$ is proportional to the DL spin-orbit effective field $H_\text{DL}$~\cite{PhysRevB.92.214406}, 
\begin{equation}
V_\text{sym} = I_\text{RF}\Delta R \mu_0 H_\text{DL} \zeta \sqrt{\frac{\mu_{0}H_\text{FMR}}{\mu_0H_\text{FMR} + \mu_{0}M_\text{eff}}}, \label{Sy}
\end{equation}
where $\zeta= ({\mu_{0}H_\text{FMR} + \mu_{0}M_\text{eff}})/[ {2{\sqrt2} W(2\mu_{0}H_\text{FMR} + \mu_{0}M_\text{eff})}]$. 
$I_\text{RF}$ and $\Delta R$ are the RF current and the resistance change due to the AMR in the bilayer, respectively. The magnitude of the antisymmetric component $V_\text{anti}$ is proportional to the sum of the FL spin-orbit effective field $H_\text{FL}$ and the Oersted field $H_\text{Oe}={j_\text{c}^\text{Pt} t_\text{Pt}}/{2}$, where $t_\text{Pt}$ is the thickness of the Pt layer and $j_\text{c}^\text{Pt}$ is the charge current density in the Pt layer~\cite{PhysRevB.92.214406}: 
\begin{equation}
V_\text{anti} = I_\text{RF}\Delta R (\mu_0 H_\text{FL}+\mu_0 H_\text{Oe}) \zeta, \label{Anti}
\end{equation}
In a system where $H_\text{FL}$ is negligible compared to $H_\text{Oe}$, the DL-SOT efficiency corresponds to the FMR spin-torque generation efficiency, defined as~\cite{liu2011spin},
\begin{equation}
\xi_\text{FMR}=\frac{{V_\text{sym}}}{{V_\text{anti}}}\frac{{e\mu_{0}M_\text{s}t_\text{Co} t_\text{Pt}}}{{\hbar}}\sqrt{1+\frac{{\mu_0{M_\text{eff}}}}{\mu_0{H_\text{FMR}}}},
\end{equation} 
where $t_\text{Pt}$ is the thickness of the Pt layer and $M_\text{s}$ is the saturation magnetization of the Co layer. However, previous studies have shown that a sizable FL-SOT is generated by the Rashba-Edelstein effect at the Co/Pt interface~\cite{PhysRevB.92.064426,PhysRevLett.116.126601}. In the presence of a non-negligible FL-SOT, the determination of the DL- and FL-SOT efficiencies, 
\begin{equation}
\xi_\text{DL(FL)} ^E = \left( \frac{2e}{\hbar } \right) \mu_{0}M_\text{s}t_\text{Co}\frac{H_\text{DL(FL)}}{E},
\end{equation} 
require to disentangle $H_\text{FL}$ and $H_\text{Oe}$ in the ST-FMR signal, where $E$ is the applied electric field. This is possible by measuring the ST-FMR with various $t_\mathrm{Co}$ because of the different $t_\text{Co}$ dependence of $H_\text{Oe}$ and $H_\text{FL}$; $H_\text{Oe}$ is independent of $t_\text{Co}$, while $H_\text{FL}$ decreases with $t_\text{Co}$. Under the assumption that $\xi_\text{DL(FL)}^E$ does not have a strong dependence on $t_\text{Co}$ in the range examined, $\xi_\text{DL(FL)}^E$ can be determined from the $t_\text{Co}$ dependence of $\xi_\mathrm{FMR}$
using~\cite{PhysRevB.92.064426}
\begin{equation}
\frac{1}{\xi_{\mathrm{FMR}}}=\frac{1}{\xi_{\mathrm{DL}}^{E}}\left(\frac{1}{\rho_{\mathrm{Pt}}}+\frac{\hbar}{e} \frac{\xi_{\mathrm{FL}}^{E}}{4 \pi M_\text{s} t_{\mathrm{Co}} t_{\mathrm{Pt}}}\right),  \label{thicknesseq}
\end{equation}
where $\rho_\mathrm{Pt}$ is the resistivity of the Pt layer.

\section{III. Molecular Tuning of Surface Rashba-Edelstein Spin-Orbit Torque}

In Figs.~\ref{fig2}(a)-\ref{fig2}(c), we show the $V_\mathrm{mix}$ signals for various $t_\mathrm{Co}$, measured using a bias tee at room temperature. We also show $1/\xi_\text{FMR}$ as a function of $1/t_\text{Co}$ for the Pt/Co bilayers in Figs.~\ref{fig2}(d)-\ref{fig2}(f). Using this result with Eq.~(\ref{thicknesseq}), we determined $\xi_\text{DL} ^E$ and $\xi_\text{FL} ^E$ for the Pt/Co, ODT-decorated Pt/Co (ODT-Pt/Co), and PFDT-decorated Pt/Co (PFDT-Pt/Co) films.

\begin{table}[]
\caption{The summarized parameters for the pristine and SAM-decorated Pt/Co films: the DL-SOT efficiency $\xi_\text{DL}^E$, the FL-SOT efficiency $\xi_\text{FL}^E$, the resistivity $\rho_\mathrm{Pt}$ of the Pt layer, and the work function $\varPhi$. The Pt resistivity $\rho_\mathrm{Pt}$ was determined from measured resistance of the pristine and SAM-decorated Pt/Co films based on a parallel circuit model with the measured resistivity of the Co layer, $\rho_\mathrm{Co}=42.1$ $\mu \Omega$cm.
}
\begin{tabular}{lccccccc}
\hline \hline
&  Pt/Co& PFDT-Pt/Co  & ODT-Pt/Co \\ \hline
$\xi_\text{DL}^E$ ($ 10^4\ \Omega^{-1}$m$^{-1}$) & $3.79\pm0.17$ & $4.33\pm0.14$ &  $3.42\pm0.19$  \\
$\xi_\text{FL}^E$ ($ 10^4\ \Omega^{-1}$m$^{-1}$) &$2.50\pm0.17$  &  $2.58\pm0.15$   &  $2.55\pm0.17$ \\
$\rho_\text{Pt}$ ($\mu\Omega$cm)  & $84.7 \pm 4.0$  & $84.9 \pm 2.3$  & $85.3 \pm 4.1$ \\
 $\varPhi$ (eV) & $4.7\pm0.2$ & $5.0\pm0.2$ & $4.4\pm 0.4$
 \\
\hline \hline
\end{tabular}
\label{tab}
\end{table}

Our finding is that only the DL-SOT is modulated by the molecular self-assembly. From the $1/t_\text{Co}$ dependence of $1/\xi_\text{FMR}$, we obtained the DL- and FL-SOT efficiencies, $\xi_\text{DL}^E$ and $\xi_\text{FL}^E$, for the Pt/Co bilayers, as shown in Table~\ref{tab}. The obtained values for the pristine Pt/Co bilayer are consistent with literature values~\cite{PhysRevLett.116.126601}. Here, we note that $\xi_\text{FL}^E$ is almost unchanged by the SAM formation as shown in Table~\ref{tab}. The change of $\xi_\text{FL}^E$ due to the molecular self-assembly is less than 3\%, which is within an experimental error. In contrast, $\xi_\text{DL}^E$ is manipulated by the molecular decoration; the PFDT formation enhances $\xi_\text{DL}^E$ by 14\%, while the ODT formation suppresses $\xi_\text{DL}^E$ by 10\%.

In the Pt/Co bilayer, the SOTs can be generated by (i) the spin Hall effect in the Pt layer, (ii) the Rashba-Edelstein effect at the Pt/Co interface, and (iii) the Rashba-Edelstein effect at the Pt surface. The spin Hall effect and surface Rashba-Edelstein effect generate the SOTs through the spin-transfer mechanism. In this picture, the ratio of the DL and FL components roughly match the ratio between the real and imaginary parts of the spin-mixing conductance, and thus the dominant component of the SOT in metallic bilayers is the DL-SOT~\cite{PhysRevB.94.104420}. In contrast, the interface Rashba-Edelstein effect primary exerts a torque on the magnetization at the interface through the exchange coupling. In this scenario, the FL-SOT is much greater than the DL-SOT, especially in the thin heavy-metal-layer limit~\cite{PhysRevB.94.104420}. Thus, in the ultrathin-Pt/Co bilayer, the DL-SOT can be attributed to the bulk spin Hall effect and surface Rashba-Edelstein effect, while the FL-SOT is dominated by the interface Rashba-Edelstein effect. The interface-dominated FL-SOT is consistent with the fact that $\xi_\mathrm{FL}^E$ is unchanged by the molecular self-assembly; because the charge screening length in Pt is less than $0.5$ $\rm\AA$, the Pt/Co interface, as well as the Co layer, cannot be affected by the ODT and PFDT formations on the Pt surface.
Here, in the Pt/Co bilayer, there is a possibility that a part of the Co layer is exposed to the surface due to a possible discontinuity of the thin Pt layer. The Rashba-Edelstein effect at the Co surface can also generate the FL-SOT~\cite{PhysRevB.96.214430}. The negligible change of $\xi_\mathrm{FL}^E$ indicates that the effect of the SAM formation on the bare Co surface is also negligible in the observed molecular tuning of the SOT.

The observed change of the DL-SOT induced by the molecular self-assembly originates from the molecular tuning of the surface Rashba-Edelstein effect in the Pt/Co bilayer. We note that the electron transport in the bulk of the Pt layer is also unaffected by the SAM formations, which is evidenced by the negligible change of the resistivity $\rho_\mathrm{Pt}$, or the carrier density, of the Pt layer after the formation of the ODT and PFDT, as shown in Table~\ref{tab}. The change of $\rho_\mathrm{Pt}$ due to the SAM formations is less than 1\%. This indicates that the observed change of $\xi_\mathrm{DL}^E$ cannot be attributed to the change of the bulk spin Hall effect, since 
the spin Hall effect can be changed through the change of the carrier density, or $\rho_\mathrm{Pt}$~\cite{dushenko2018tunable}. 
Since the spin Hall effect, as well as the interface Rashba-Edelstein effect, is unaffected by the SAM formations, the change of $\xi_\mathrm{DL}^E$ can only be attributed to the change of the DL-SOT arising from the Rashba-Edelstein effect at the Pt surface, tuned by the molecular self-assembly.

\begin{figure}[tb]
\includegraphics[scale=1]{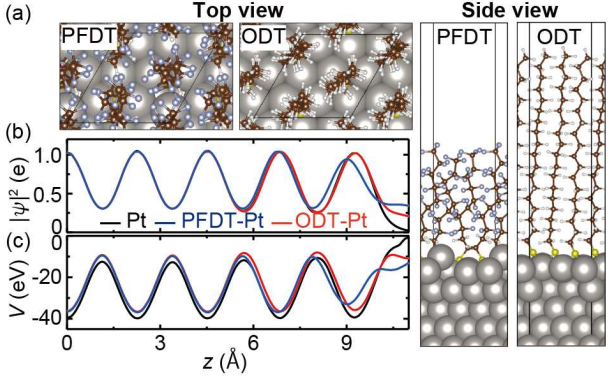}
\caption{
(a) Top view and side view of optimized geometries of PFDT- and ODT-adsorbed surfaces. Brown, yellow, blue, white, and gray spheres represent C, S, F, H, and Pt atoms, respectively. Unit cell is shown by the solid line. (b) Plane averaged charge density $|\psi|^2$ and (c) electric potential $V$ for Pt (black), PFDT-Pt (blue) and ODT-Pt (red). Coordinate represents the $z$-axis.} 

\label{fig3} 
\end{figure}

The observed modulation of the SOT induced by the molecular self-assembly indicates that the surface Rashba-Edelstein generates a sizable SOT in the ultrathin-Pt/Co bilayers. Under the assumption that the DL-SOT is generated by the spin-transfer mechanism, the measured $\xi_\text{DL}^E$ can be decomposed into two terms: $\xi_\text{DL}^E=\xi_\text{DL,bulk}^E+\xi_\text{DL,surface}^E$, where $\xi_\text{DL,bulk}^E$ and $\xi_\text{DL,surface}^E$ are the DL-SOT efficiency due to the bulk spin Hall effect and surface Rashba-Edelstein effect, respectively. The DL-SOT efficiency due to the spin Hall effect is expressed as~\cite{PhysRevLett.116.126601}
\begin{equation}
\xi_{\mathrm{DL,bulk}}^{E}=\frac{2 e}{\hbar} \sigma_{\mathrm{SH}}\left[1-\operatorname{sech}\left(\frac{t_{\mathrm{Pt}} }{ \lambda_\text{Pt}}\right)\right]\left[1+\frac{\sigma_{\mathrm{Pt}}\tanh \left(t_{\mathrm{Pt}} / \lambda_\text{Pt}\right)}{2 \lambda_\text{Pt}  G_\text{r}}\right]^{-1}, \label{xiEbulk}
\end{equation}
where $\sigma_\mathrm{SH}$, $\sigma_\mathrm{Pt}$, and $\lambda_\mathrm{Pt}$ are the spin Hall conductivity, conductivity, and spin diffusion length of the Pt layer, respectively. $G_\mathrm{r}$ is the real part of the spin-mixing conductance. Using Eq.~(\ref{xiEbulk}) with $G_\mathrm{r}= 0.59\times10^{15}$ $\Omega^{-1}$m$^{-1}$, $\sigma_\mathrm{SH}=1.8\times10^5$ $\Omega^{-1}$m$^{-1}$ and $\lambda_\mathrm{Pt}/\sigma_\mathrm{Pt}=0.77\times 10^{-15}$ $\Omega$m$^2$~\cite{PhysRevLett.116.126601,sagasta2016tuning}, we obtain $\xi_{\mathrm{DL,bulk}}^{E}=2.42\times10^{4}$ $\Omega^{-1}$m$^{-1}$.
For the ultrathin-Pt/Co bilayer, from the measured $\xi_\mathrm{DL}^E$, we obtain the DL-SOT efficiency due to the surface Rashba-Edelstein effect in the Pt/Co bilayer as $\xi_\text{DL,surface}^E=\xi_\text{DL}^E-\xi_\text{DL,bulk}^E=1.37\times 10^4$ $\Omega^{-1}$m$^{-1}$. Here, $\xi_\text{DL,surface}^E$ is proportional to the conversion efficiency from the applied 2D charge current $j_\mathrm{c}^\text{2D}$ into the 3D spin current $j_\mathrm{s}^\mathrm{3D}$, $q_\mathrm{REE}=j_\mathrm{s}^\mathrm{3D}/j_\mathrm{c}^\text{2D}$, at the Pt surface
as~\cite{kondou2018efficient} 
\begin{equation}
\mathop \xi \nolimits_{{\rm{DL,surface}}}^E  = \frac{{\left( {2\lambda_\mathrm{Pt} /\sigma_\mathrm{Pt} } \right){G_{\rm{r}}}{\mathop{\rm sech}\nolimits} \left( {{t_{{\rm{Pt}}}}/{\lambda _{{\rm{Pt}}}}} \right)}}{{1 + \left( {2\lambda_\mathrm{Pt} /\sigma_\mathrm{Pt} } \right){G_{\rm{r}}}\tanh \left( {{t_{{\rm{Pt}}}}/{\lambda _{{\rm{Pt}}}}} \right)}}\frac{{\sigma_{\rm s}{\lambda _{{\rm{Pt}}}}{q_{{\rm{REE}}}}}}{{2\sqrt 2 }}, \label{xiEsurface}
\end{equation}
where $\sigma_\text{s}$ is the conductivity of the Pt surface. 
Assuming $\sigma_\text{s}=\sigma_\mathrm{Pt}$, 
we obtain $q_\mathrm{REE} =0.045$ nm$^{-1}$ for the surface Rashba state of Pt. This value is comparable to $q_\mathrm{REE} $ for other Rashba systems, such as the interfaces between a nonmagnetic metal and indium-tin-oxide~\cite{kondou2018efficient}.
Using the obtained parameters, we also calculated Pt-layer thickness $t_\mathrm{Pt}$ dependence of $\xi_\text{DL,bulk}^E$, $\xi_\text{DL,surface}^E$, and $\beta=\xi_\text{DL,bulk}^E/(\xi_\text{DL,bulk}^E+\xi_\text{DL,surface}^E)$ for Pt/Co bilayers as shown in Figs.~\ref{fig2}(g) and \ref{fig2}(h).
For the calculation, we have taken into account $t_\mathrm{Pt}$ dependence of the Pt-layer resistivity $\rho_\mathrm{Pt}$ [see the inset to Fig.~\ref{fig2}(h)]~\cite{doi:10.1063/1.1644906}.

Figures~\ref{fig2}(g) and \ref{fig2}(h) demonstrate that the surface contribution is significant only when $t_\mathrm{Pt}<3$ nm. In fact, we have confirmed that $\xi_\mathrm{DL}^E$ measured for a Pt/Co bilayer with $t_\mathrm{Pt}=10$ nm is not affected by the SAM formations. The negligible contribution from $\xi_\text{DL,surface}^E$ to $\xi_\text{DL}^E$ when $t_\mathrm{Pt}>3$ nm is consistent with previous reports where the surface effect has been neglected. 

From the measured values of $\xi_\text{DL}^E$ shown in Table~\ref{tab}, we obtained 
$\xi_\text{DL,surface}^E=1.91\times 10^4$ $\Omega^{-1}$m$^{-1}$ for the PFDT-Pt/Co, and $\xi_\text{DL,surface}^E=1.00\times 10^4$ $\Omega^{-1}$m$^{-1}$ for the ODT-Pt/Co under the assumption that the $\xi_\text{DL,bulk}^E$ is unchanged by the molecular self-assembly, which is supported by the negligible change of $\rho_\mathrm{Pt}$. This result shows that the PFDT formation enhances the surface Rashba-Edelstein SOT efficiency $\xi_\text{DL,surface}^E$ by 39\%, while the ODT formation suppresses $\xi_\text{DL,surface}^E$ by 27\%. 
The enhanced(suppressed) surface Rashba effect induced by the PFDT(ODT) formation is supported by spin pumping experiments~\cite{Supplemental}.

\section{IV. density functional theory calculations}

To reveal the mechanism of the molecular tuning of the surface Rashba-Edelstein SOT, we performed the plane-wave based density functional theory (DFT) calculations with VASP program package~\cite{PhysRevB.54.11169}. We assumed the ODT- and PFDT-adsorbed surface structure as  ($\sqrt{3}\times\sqrt{3})R30^{\circ}$ unit cells [see Fig.~\ref{fig3} and Ref.~\cite{Supplemental}]. This surface structure is often found in the thiol-adsorbed
Pt surfaces~\cite{C4RA04659E}. The BEEF-vdW functional was used~\cite{PhysRevB.85.235149}, and the plane wave cutoff was set to 400 and 500 eV for the geometry optimization and electrostatic potential calculations, respectively. We placed $\sim15$ {\AA} vacuum layers between surface slabs, and $6\times 6\times 1$ $k$-point was used for the reciprocal space integration.

The calculation reveals that the manipulation of the surface Rashba-Edelstein effect is induced by molecular tuning of the wavefunction asymmetry and electrostatic potential near the Pt surface. At metal surfaces, the symmetry-breaking surface potential distorts the atomic orbitals through the admixture with the other orbitals, giving rise to asymmetric wavefunctions close to the nucleus, where the spin-orbit coupling is strong~\cite{PhysRevB.90.165108,petersen2000simple,bihlmayer2006rashba,PhysRevB.71.201403}. The wavefunction asymmetry 
plays a key role in the Rashba physics~\cite{petersen2000simple}. In fact, tight-binding calculations have shown that the strength of the Rashba spin-orbit coupling is determined by the asymmetry of the surface state wavefunction near the position of the nucleus, in combination with the atomic spin-orbit coupling~\cite{bihlmayer2006rashba}. In Fig.~\ref{fig3}(a), we show the optimized geometry of the PFDT- and ODT-adsorbed surfaces. For the optimized structure, we calculated the planar averaged charge density and the planar averaged electrostatic potential as shown in Figs.~\ref{fig3}(b) and \ref{fig3}(c). 
This result shows that both the change density and electrostatic potential close to the Pt surface is modulated by the SAM formation.  

In plane-wave-based calculations, the Rashba parameter can be expressed as~\cite{Nagano_2009}
\begin{equation}
\alpha_{\rm R} = \frac{\hbar^2}{4m^2 c^2}\int \frac{\partial V }{\partial z }|\psi|^2  \ dz, \label{rashbaeq}
\end{equation}
where $c$, $\partial V/\partial z$ and $|\psi|^2$  are the speed of light, potential gradient, and electron density distribution, respectively. Equation~(\ref{rashbaeq}) indicates that the asymmetry of the charge density distribution and potential gradient gives rise to the Rashba spin-orbit coupling. 
Using Eq.~(\ref{rashbaeq}) with the charge density distribution and electrostatic potential shown in Figs.~\ref{fig3}(b) and \ref{fig3}(c), we calculated the Rashba parameter. The calculation shows that the Rashba parameter $\alpha_{\rm R}^{\rm Pt}$ of the Pt surface is enhanced by the PFDT formation, while that is decreased by the ODT formation: $\alpha_{\rm R}^{\rm PFDT-Pt} / \alpha_{\rm R}^{\rm Pt}=1.4$ and $\alpha_{\rm R}^{\rm ODT-Pt} / \alpha_{\rm R}^{\rm Pt}= 0.94$, where $\alpha_{\rm R}^{\rm PFDT(ODT)-Pt}$ is the Rashba parameter of the PFDT(ODT)-Pt. Although the ODT-induced change of $\alpha_\mathrm{R}$ in this simple model is smaller than the experimental value, this result is qualitatively consistent with the enhanced(suppressed) surface Rashba-Edelstein SOT induced by the PFDT(ODT) formation.

The calculation shows that both PFDT and ODT formations on the Pt surface induce the transfer of electrons from the Pt to the molecule; 0.320 electrons per Pt atom are transferred to PFDT and 0.268 electrons per Pt atom are transferred to ODT. This indicates that the electron transfer alone cannot explain the fact that the PFDT formation enhances $\alpha_\mathrm{R}$, while the ODT formation suppresses $\alpha_\mathrm{R}$. We note that the change of $\alpha_\mathrm{R}$ due to the PFDT formation is strongly affected by the out-of-plane buckling of Pt atoms; as shown in Fig.~\ref{fig3}(a), a Pt atom of the outermost surface adsorbed by PFDT is protruding upward while other atoms are pushed into the bulk. Such surface reconstruction is insignificant in ODT-Pt. This surface reconstruction can be conjectured to optimize the hybridization of the ($s$, $p_{\rm z}$) surface state of PFDT-Pt. The buckling of surface atoms has been shown to play an important role to realize strong spin splitting, such as in heavy metal/Ag(111) and 2D honeycomb systems~\cite{PhysRevB.75.195414, PhysRevB.96.161401}. The interface phenomena manipulated by molecular layers have been investigated in the field of spinterface~\cite{Spinterface, Cinchetti2017}. The research in this field mainly focuses on the manipulation of magnetic properties of ferromagnetic metals, such as the magnetic anisotropy and spin polarization. Our result demonstrates that organic molecules also provide a way to control the SOTs in metallic heterostructures, extending the research field of spinterface and spin-orbitronics.

\section{V. CONCLUSIONS}

In summary, we demonstrated that the surface Rashba-Edelstein effect 
generates a sizable SOT. 
We note that the surface Rashba-Edelstein SOT had not been observed even in Pt, the most fundamental and widely used heavy metal in spin-orbitronics. 
We also note that despite a theoretical prediction of a strong Rashba effect~\cite{PhysRevB.83.195427}, the observation of the Rashba spin-splitting at Pt surfaces using the photoelectron spectroscopy is still lacking. 
Our discovery of the surface Rashba-Edelstein SOT in the prototypical spin-orbitronic system alters the landscape of the SOT generation, promising important advances in the spin-orbit physics of solid-state devices.

\begin{acknowledgments}
This work was supported by JSPS KAKENHI Grant Numbers 19H00864, 26220604, 26103004, 18J20780, 17H04808, Asahi Glass Foundation, Kao Foundation for Arts and Sciences, Canon Foundation, JGC-S Scholarship Foundation, and Spintronics Research Network of Japan (Spin-RNJ). 
\end{acknowledgments}

%

\end{document}